\documentclass[english,aps,twocolumn]{revtex4}

\usepackage[latin1]{inputenc}
\usepackage{graphicx}
\usepackage{amssymb}

\makeatletter

\usepackage{epsfig}
\usepackage{color}
\usepackage{ifthen}

\newcommand{\bra}[1]{{\langle #1 |}}
\newcommand{\ket}[1]{{| #1 \rangle}}
\newcommand{\expect}[1]{{\left\langle #1 \right\rangle}}
\newcommand{\mubohr}{{\mu_{\rm B}}}

\newcommand{\spup}{\ket{\!\uparrow}}
\newcommand{\spdown}{\ket{\!\downarrow}}

\newcommand{\rhoStat}{\bar{\rho}}
\newcommand{\rhoSysStat}{\bar{\rho}_\sysLbl}
\newcommand{\rhoSys}{\rho_\sysLbl}
\newcommand{\rhoDotSys}{\dot{\rho}_\sysLbl}

\newcommand{\rSNot}{\rhoSys^0}
\newcommand{\rBNot}{{\rho_\bathLbl^0}}

\newcommand{\sopst}[1]{{\cal #1}}   %sop style
\newcommand{\op}[2]{\ket{#1}\bra{#2}}
\newcommand{\tunnelInDotLbl}{\mathrm{in}}
\newcommand{\tunnelOutDotLbl}{\mathrm{out}}

\newcommand{\fermiUCutoff}{{\epsilon_c}}

\newcommand{\expectSopTemplateThree}[3]{#3_{#1}\ifthenelse{\equal{#2}{}}{}{^{#2}}}  %={#3_{#1}^{#2}}
\newcommand{\expectSopSymbol}{\sopst{W}}
\newcommand{\expectSopTemplate}[2]{\expectSopTemplateThree{#1}{#2}{\expectSopSymbol}}
\newcommand{\expectSopHCTemplate}[2]{\expectSopTemplateThree{#1}{#2}{\applySopAndHC{\expectSopSymbol}}}
\newcommand{\expectSop}[1]{\expectSopTemplate{#1}{>}}

\newcommand{\expectSopTwo}[2]{\expectSopTemplate{#1,#2}{}}
\newcommand{\expectSopTwoHC}[2]{\expectSopHCTemplate{#1,#2}{}}

\newcommand{\expectSopInt}[1]{\expectSopTemplate{#1}{>,\int}}
\newcommand{\expectSopIntAndHC}[1]{\expectSopHCTemplate{#1}{>,\int}}

\newcommand{\expectSopRight}[1]{\expectSopTemplate{#1}{<}}
\newcommand{\expectSopRightAndHC}[1]{\expectSopHCTemplate{#1}{<}}

\newcommand{\expectSopFull}[1]{\expectSopTemplate{#1}{\rm f}}
\newcommand{\expectSopFullInt}[1]{\expectSopTemplate{#1}{{\rm f}\int}}
\newcommand{\MeqSop}{\sopst{W}^{\rm meq}}

\newcommand{\applySopAndHC}[1]{\widetilde{#1}}
\newcommand{\applySopSymm}[1]{\overline{#1}}
\newcommand{\applySopAntisymm}[1]{{#1}''}

\newcommand{\TrB}{\mathrm{Tr}_\bathLbl \,}
\newcommand{\TrS}{\mathrm{Tr}_\sysLbl \,}
\newcommand{\TrF}{\mathrm{Tr}\,}
\newcommand{\LS}{{L_\sysLbl}}

\newcommand{\LNot}{L_0}
\newcommand{\LV}{{L_V}}

\newcommand{\sopIx}[4]{\ifthenelse{\equal{#1#2#3#4}{}}{}{#1#2|#3#4}}  %= {#1#2|#3#4}
\newcommand{\calGSymbol}{\mathcal{G}}
\newcommand{\calG}[5]{{\calGSymbol}_{\sopIx{#1}{#2}{#3}{#4}}^{#5}}

\newcommand{\memoryKernel}{M}
\newcommand{\condRho}{\rho^{\rm c}}

\newcommand{\sqrtGammaLSSymbol}{\Gamma}
\newcommand{\sqrtGammaLS}[1]{\sqrtGammaLSSymbol_{ls,\, {#1}}}
\newcommand{\gammaTZero}{\gamma}
\newcommand{\gammaTZeroAvg}{\gamma}

\newcommand{\gl}[1]{\gammaTZero_{#1}}
\newcommand{\gls}[2]{\gammaTZero^{#2}_{#1}}
\newcommand{\gul}[1]{\gammaTZero^\uparrow_{#1}}
\newcommand{\gdl}[1]{\gammaTZero^\downarrow_{#1}}

\newcommand{\guO}{\gul{1}}
\newcommand{\guT}{\gul{2}}

\newcommand{\gO}{\gammaTZero_1}
\newcommand{\gT}{\gammaTZero_2}
\newcommand{\gAvg}{\gammaTZeroAvg}  % (\gO+\gT)/2
\newcommand{\Dm}{\Delta\mu}

\renewcommand{\expectSopFullInt}[1]{\expectSopTemplate{#1}{\mathrm{f}}}

\renewcommand{\expectSopInt}[1]{\expectSopTemplate{#1}{>}}
\renewcommand{\expectSopIntAndHC}[1]{\expectSopHCTemplate{#1}{>}}
\renewcommand{\MeqSop}{\sopst{M}}
\renewcommand{\applySopSymm}[1]{{#1}'}

\renewcommand{\LV}{L_T}
\renewcommand{\expectSopSymbol}{W}
\renewcommand{\calGSymbol}{G}

\renewcommand{\memoryKernel}{\hat{M}}

\renewcommand{\marginpar}[1]{}

\newcommand{\mycomment}[1]{}

\usepackage{babel}
\makeatother
\begin{document}

\newcommand{\Hlead}{H_{\mathrm{lead}}}

\newcommand{\Hdot}{H_{\mathrm{d}}}

\newcommand{\tauc}{\tau_{\mathrm{c}}}

\newcommand{\bathLbl}{\mathrm{R}}

\newcommand{\sysLbl}{S}

\newcommand{\correl}[1]{C_{#1}}

\newcommand{\correlInitTime}{t'}

\newcommand{\correlTimeDiff}{t}
 
\newcommand{\Wns}[1]{#1'}

\newcommand{\expectSopI}[1]{\expectSopFullInt{#1}}

\newcommand{\expectSopTwoSymmI}[2]{\applySopSymm{\expectSopTwo{#1}{#2}}}

\newcommand{\expectSopTwoAntisymmI}[2]{\applySopAntisymm{\expectSopTwo{#1}{#2}}}

\newcommand{\lalt}{l'}

\newcommand{\Ssymm}{S^{\mathrm{sym}}}

\newcommand{\calGPlusOutPMIn}{\calGSymbol_{\pm}^{l}}

\newcommand{\LindGPlusOutPMIn}{\LindSopToOpSymbol_{\pm}^{l}}

\newcommand{\calGPlusOutPlusIn}{\calGSymbol_{+}^{l}}

\newcommand{\LindGPlusOutPlusIn}{\LindSopToOpSymbol_{+}^{l}}

\newcommand{\calGMinusOutPlusIn}{\applySopAndHC{\calGSymbol}_{+}^{l}}

\newcommand{\calGPlusOutMinusIn}{\calGSymbol_{-}^{l}}

\newcommand{\LindGPlusOutMinusIn}{\LindSopToOpSymbol_{-}^{l}}

\newcommand{\calGMinusOutMinusIn}{\applySopAndHC{\calGSymbol}_{-}^{l}}

\newcommand{\lblTwoStates}{\mathrm{a}}

\newcommand{\lblThreeStatesSmZeeman}{\mathrm{b}}

\newcommand{\lblThreeStatesImZeeman}{\mathrm{c}}

\newcommand{\lblMarkov}{\mathrm{Mkv}}

\newcommand{\omegans}{\omega}

\newcommand{\DEnoise}[1]{E_{#1}}

\newcommand{\DEup}{\DEnoise{\uparrow}}

\newcommand{\DEdown}{\DEnoise{\downarrow}}

\newcommand{\DEupdown}{\DEnoise{\uparrow,\downarrow}}

\newcommand{\principalValueSym}[2]{p_{#1}^{#2}}

\newcommand{\unitySymbol}{1\hspace{-.35em}1}

\newcommand{\LindSopToOpSymbol}{g}

\newcommand{\levelWidth}{\gAvg}

\newcommand{\dotEnergy}{\DEup}

\newcommand{\leadEnergySymbol}{\epsilon}

\newcommand{\leadEnergy}{\leadEnergySymbol_{lk}}

\newcommand{\leadEnergyAlt}{\leadEnergySymbol_{l'k'}}

\newcommand{\leadAnSymbol}{c}

\newcommand{\leadAn}{\leadAnSymbol_{lk}}

\newcommand{\leadAnAlt}{\leadAnSymbol_{l'k'}}

\newcommand{\kTwo}{k''}

\newcommand{\leadAnAltTwo}{c_{l''\kTwo}}

\newcommand{\tunnelAmp}{t_{l}}

\newcommand{\biResFactor}[1]{j_{#1}}

\renewcommand*{\applySopAndHC}[1]{\tilde{#1}}

\renewcommand*{\sqrtGammaLSSymbol}{t}

\renewcommand*{\sqrtGammaLS}[1]{\sqrtGammaLSSymbol^{l\sigma}_{#1}}

\renewcommand*{\condRho}{\rho^{\rm c}_\sysLbl}

\renewcommand*{\memoryKernel}{M}

%\author{Hans-Andreas \surname{Engel}} \email{Hans-A.Engel@unibas.ch} 

%\author{Daniel \surname{Loss}}\email{Daniel.Loss@unibas.ch} 

\affiliation{Department of Physics and Astronomy, University of Basel,  Klingelbergstrasse 82, CH-4056 Basel, Switzerland}

\title{Asymmetric Quantum Shot Noise in Quantum Dots}

\author{Hans-Andreas Engel}

\author{Daniel Loss}

\affiliation{Department of Physics and Astronomy, University of Basel, Klingelbergstrasse
82, CH-4056 Basel, Switzerland}

\begin{abstract}
We analyze the frequency-dependent noise of a current through a quantum
dot which is coupled to Fermi leads and which is in the Coulomb blockade
regime. We show that the asymmetric shot noise as function of frequency
shows steps and becomes super-Poissonian. This provides experimental
access to the quantum fluctuations of the current. We present an exact
calculation for a single dot level and a perturbative evaluation of
the noise in Born approximation (sequential tunneling regime but without
Markov approximation) for the general case of many levels with charging
interaction. 
\end{abstract}
\maketitle
The shot noise is a striking consequence of charge quantization and
allows to characterize the transport of individual electrons \cite{BlanterBuettiker}.
 The symmetry of the noise $S(\omegans)$ is important: For a classical
stationary system, the noise (for autocorrelations) is always symmetric
in the frequency $\omegans$. However, for a quantum system, the noise
can by asymmetric due to the non-commutativity of current operators
at different times. It was recently found that such an asymmetric
noise can be detected  since the noise frequency $\omega$ corresponds
to a quantum of energy $\hbar\omega$ which is transferred from the
system to the measurement apparatus \cite{Lesovik,Aguado,Imry} which
was demonstrated experimentally \cite{DeblockNoiseMeas}. This means
that antisymmetric quantum effects in the noise can be measured and
isolated from the classical (symmetric) effects.
 We show that striking asymmetric effects appear in the shot noise
$S(\omega)$ of a quantum dot with steps as function of $\omegans$,
giving a super-Poissonian Fano factor.  Our analysis is based on
a perturbative approach which remains valid in the quantum limit with
large $\omegans$ (where a Markov approximation typically invoked
would not be valid). We confirm our perturbative results by an exact
calculation of the noise for a dot with a single level. We note that
quantum dots are good candidates for an experimental test of our predictions
since such systems have been studied extensively over the years, both
experimentally and theoretically \cite{Birk,Averin93,Korotkov94,Hershfield,DDot,NoiseLong}.

We consider the operator $I_{l}$ which describes the current in a
lead $l$. We define the current noise, \begin{equation}
S_{l\lalt}(\omegans)=\int_{-\infty}^{\infty}\! dt\: e^{i\omegans\correlTimeDiff}\left[\expect{I_{l}(\correlTimeDiff)I_{\lalt}}-\expect{I_{l}}\expect{I_{\lalt}}\right],\label{eqnNoiseDef}\end{equation}
in terms of the (non-symmetrized) correlation function, $\expect{I_{l}(\correlTimeDiff)I_{\lalt}}=\TrF I_{l}(\correlTimeDiff)I_{\lalt}\rhoStat$.
Here, $\rhoStat$ is the stationary density matrix (of the full quantum
system). Note that $I_{l}(\correlTimeDiff)I_{\lalt}$ is not a Hermitian
operator and thus does not correspond to a classical observable. How
should we interpret Eq.~(\ref{eqnNoiseDef})? On the one hand, one
can avoid  the non-Hermitian operator by arguing heuristically that
Eq.\ (\ref{eqnNoiseDef}) is {}``unphysical'' and by instead considering
the correlation function in terms of the symmetrized operator $\frac{1}{2}\left[I_{l}(\correlTimeDiff)I_{\lalt}+I_{\lalt}I_{l}(\correlTimeDiff)\right]$
\cite{LandauLifshitzS118}. One then obtains the symmetrized noise,
$\Ssymm_{l\lalt}(\omegans)=\frac{1}{2}\left[S_{l\lalt}(\omegans)+S_{\lalt l}(-\omegans)\right]$.
For $l=\lalt$, this noise does not depend on the sign of $\omegans$.
This corresponds to a measurement apparatus \cite{Lesovik,Aguado,Imry}
which does not discriminate between absorption and emission of energy
by the system, and thus cannot detect all noise characteristics.
On the other hand, we have $\expect{I_{l}(-\correlTimeDiff)I_{\lalt}}=\expect{I_{\lalt}(\correlTimeDiff)I_{l}}^{*}$
and thus $S_{l\lalt}(\omegans)=S_{\lalt l}(\omegans)^{*}$, since
$I_{l}$ is Hermitian and $\rhoStat$ stationary. Thus, $S_{ll}(\omegans)$
is a real quantity which can be regarded as an observable. Indeed,
there are setups, where $S_{ll}(\omegans)$ can be accessed experimentally
\cite{Lesovik,Aguado,Imry,DeblockNoiseMeas}. This justifies to consider
the asymmetric quantum noise {[}Eq.\ (\ref{eqnNoiseDef}){]} as we
do in the following.

\emph{Quantum dots}. To illustrate the presence of asymmetric shot
noise contributions due to quantum effects, we consider now a concrete
system of a quantum dot in the Coulomb blockade regime \cite{kouwenhoven},
coupled to Fermi leads $l=1,2,...$ at chemical potentials $\mu_{l}$.
When only a single dot level is present, the noise can be calculated
even exactly \cite{Averin93} (see below). This is however not possible
for systems with many levels and charging interaction, for which we
now develop a perturbative approach. We assume weak coupling such
that current and noise are dominated by the sequential tunneling (ST)
contributions, valid for $kT>\gAvg$ with temperature $T$ and level
width $\gAvg$. We model the combined system with the Hamiltonian
$H=\Hlead+\Hdot+H_{T},$ which describes leads, dot, and the tunnel
coupling between leads and dot, resp., and with $H_{0}=\Hlead+\Hdot$.
We let $\Hlead=\sum_{lk\sigma}\epsilon_{lk}c_{lk\sigma}^{\dagger}c_{lk\sigma}$,
where $c_{lk\sigma}^{\dagger}$ creates an electron in lead $l$ with
orbital state $k$, spin $\sigma$, and energy $\epsilon_{lk}$. The
electronic dot states $\ket{n}$ are described by $\Hdot\ket{n}=E_{n}\ket{n}$,
including charging and interaction energies. We use the standard tunneling
Hamiltonian $H_{T}=\sum_{lpk\sigma}t_{lp}^{\sigma}c_{lk\sigma}^{\dagger}d_{p\sigma}+\textrm{H.c.},$
with tunneling amplitude $t_{lp}^{\sigma}$ and where $d_{p\sigma}^{\dagger}$
creates an electron on the dot with orbital state $p$ and spin $\sigma$.
The state of the combined system is given by the full density matrix
$\rho$, while the electronic states of the dot are described by the
reduced density matrix, $\rhoSys=\TrB\rho$, where the trace is taken
over the leads. We assume that at some initial time $t_{0}$ the full
density matrix factorizes, $\rho(t_{0})=\rSNot\rBNot$, with the equilibrium
density matrix of the leads, $\rBNot$. From the von Neumann equation
$\dot{\rho}=-i\,[H,\,\rho]$  one finds \cite{FickSauermann} the
generalized master equation for the reduced density matrix, $\rhoDotSys(t)=-i\,\LS\rhoSys(t)-\int_{t_{0}}^{t}dt'\memoryKernel(t')\rhoSys(t\!-\! t')$,
where the kernel $\memoryKernel$ is the self-energy superoperator.
Since we consider the weak coupling regime, we proceed with a systematic
lowest-order expansion in $H_{T}$. We obtain $\memoryKernel(\tau)=\TrB\LV\, e^{-i\LNot\tau}\LV\rBNot$,
where we define the superoperators $\LS X=\left[\Hdot,\, X\right]$,
$\LNot X=\left[H_{0},\, X\right]$, and $\LV X=\left[H_{T},\, X\right]$.
In the following, we work in the Laplace space, $f(t)\mapsto f(\omega)=\int_{0}^{\infty}dt\: e^{i\omega t}f(t)$
(we take $\mathrm{Im}\,\omega>0$ but our results remain well-defined
for $\mathrm{Im}\,\omega\to0$). Then, the time evolution of $\rhoSys$
reads\begin{equation}
-\rhoSys(t_{0})-i\omega\rhoSys(\omega)=\MeqSop(\omega)\:\rhoSys(\omega),\label{eqnMasterEq}\end{equation}
with $\MeqSop(\omega)=-i\LS-\memoryKernel(\omega)$ and with the lower
boundary of the Laplace transformation shifted to $t_{0}$. We take
$t_{0}\to-\infty$ and assume that the system has relaxed at the much
later time $t=0$ into its stationary state $\rhoSysStat=\rhoSys(0)=\lim_{\omega\to0}(-i\omega)\rhoSys(\omega)$.
 We multiply Eq.\ (\ref{eqnMasterEq}) by $-i\omega$, take $\omega\to0$,
and find the equation $\MeqSop(0)\,\rhoSysStat=0$ from which we get
$\rhoSysStat$.

${}$%
\marginpar{current superoperators%
}\emph{Current}. We calculate the current $I_{l}$ flowing from the
dot into lead $l$ and vice versa. The current operators are $I_{l}(t)=(-1)^{l}e\dot{q}_{l}(t)=(-1)^{l}ie\left[H_{T},\, q_{l}(t)\right]$
where $q_{l}=\sum_{k\sigma}c_{lk\sigma}^{\dagger}c_{lk\sigma}$ is
the number of electrons in lead $l$. We choose the sign of $I_{l}$
such that $\expect{I_{1}}=\expect{I_{2}}$ in the case of two leads.
We now introduce the projectors $P=\rBNot\TrB\!$ and $Q=\unitySymbol-P$
with the properties $P\LV P=0=PI_{l}P$ and $P\LNot=\LNot P$. We
evaluate $\expect{I_{l}}$  by inserting $P+Q=\unitySymbol$ and
find $\expect{I_{l}(t)}=\TrF I_{l}Qe^{-i(\LNot+\LV)(t-t_{0})}\rho(t_{0})=-i\,\TrF I_{l}\int_{t_{0}}^{t}dt'\: Qe^{-i\LNot(t-t')}\LV P\rho(t')+O(H_{T}^{3}).$
This motivates introducing the following superoperators, $\expectSopFull{l}=\expectSop{l}+\expectSopRight{l}$
with $\expectSop{l}\left(\tau\right)=-i\,\TrB I_{l}\, e^{-i\LNot\,\tau}\LV\,\rBNot$
and $\expectSopRight{l}\left(\tau\right)=-i\,\TrB\LV e^{-i\LNot\,\tau}I_{l}\,\rBNot$,
and $\expectSopTwo{l}{\lalt}(\tau)=\TrB I_{l}\, e^{-iL_{0}\tau}I_{\lalt}\rBNot$.
Note that these superoperators act only on the dot space, which considerably
simplifies further evaluations. %
\marginpar{evaluate current%
} In the ST regime, the current is\begin{equation}
\expect{I_{l}}=\TrS\expectSopFull{l}(\omega\!=\!0)\,\rhoSysStat.\label{eqnCurrent}\end{equation}
 This implies that the superoperator $\expectSopFull{l}$ corresponds
to the current. 

\emph{Quantum shot noise}. We evaluate the noise {[}Eq.\ (\ref{eqnNoiseDef}){]}
in lowest order in $H_{T}$ but without any further approximation.
It is sufficient to consider $\correlTimeDiff>0$, since $\expect{I_{l}(-\correlTimeDiff)I_{\lalt}}=\expect{I_{\lalt}(\correlTimeDiff)I_{l}}^{*}$.
Using again $P+Q=\unitySymbol$ we get $\expect{I_{l}(\correlTimeDiff)\, I_{\lalt}}=\TrF I_{l}Qe^{-iL\correlTimeDiff}PI_{\lalt}Q\rhoStat+\TrF I_{l}Qe^{-iL\correlTimeDiff}QI_{\lalt}P\,\rhoStat$,
while we neglect the higher-order term $\TrF I_{l}Qe^{-iLt}QI_{l'}Q\,\rhoStat$
in $H_{T}$. The goal is to factor out one of the following expressions.
First, we expand $Qe^{-iLQt}Q=Qe^{-i\LNot t}Q+O(H_{T})$ to leading
order. Second, we evaluate the conditional time evolution $\condRho(t):=\TrB e^{-iLt}\rBNot$.
We find that $\condRho$ is the formal solution of Eq.\ (\ref{eqnMasterEq})
with initial value $\unitySymbol$, thus $\condRho(\omega)=-[i\omega+\MeqSop(\omega)]^{-1}$
\cite{condRhoResum}. Finally, we use $\TrB I_{l}Q\, e^{-iLt}\rBNot=\int_{0}^{t}dt'\:\expectSop{l}\left(t-t'\right)\,\condRho\left(t'\right)$
and $\TrB I_{l}Qe^{-iLt}QI_{\lalt}\rBNot=\expectSopTwo{l}{\lalt}\left(t\right)+\int_{0}^{t}dt'\int_{0}^{t'}\! dt''\:\expectSop{l}\left(t-t'\right)\:\condRho\left(t'-t''\right)\:\expectSopRight{\lalt}\left(t''\right)$.
We obtain the noise correlation in the ST regime,\begin{eqnarray}
\lefteqn{S_{l\lalt}(\omegans)=2\TrS\Big\{\expectSopFull{l}(\omega)\,\condRho(\omega)\left[\expectSopInt{\lalt}(0)+\expectSopRight{\lalt}\left(\omega\right)\right]+}\nonumber \\
 & + & \expectSopFull{\lalt}(-\omega)\,\condRho(-\omega)\left[\expectSopIntAndHC{l}(0)+\expectSopRightAndHC{l}\left(-\omega\right)\right]\nonumber \\
 & + & \expectSopTwo{l}{l'}(\omega)+\expectSopTwoHC{l'}{l}(-\omega)\Big\}\rhoSysStat.\label{SgeneralW}\end{eqnarray}
Here, $\omega$ is real and the limit $\omega\to0$ is well behaved
{[}the $\delta(\omegans)$ contribution is cancelled by $\expect{I_{l}}\expect{I_{\lalt}}$
in Eq. (\ref{eqnNoiseDef}){]}. For a superoperator $\sopst{S}$,
we have defined $\applySopAndHC{\sopst{S}}$ such that $(\sopst{S}A)^{\dagger}=\applySopAndHC{\sopst{S}}A^{\dagger}$,
thus $\sopst{S}_{bc|nm}:=\bra{b}\big(\sopst{S}\op{n}{m}\big)\ket{c}=(\applySopAndHC{\sopst{S}}_{cb|mn})^{*}$,
and $\applySopAndHC{\sopst{S}}(\omega)A=\int_{0}^{\infty}\!\! dt\, e^{i\omega t}[\sopst{S}(t)A]^{\dagger}$.
For deriving Eqs.\ (\ref{eqnCurrent})-(\ref{SgeneralW}), we have
made no Markov approximation where we would evaluate $\memoryKernel(t)e^{it\LNot}$
at $\omega=0$ and equivalently for the other superoperators \cite{MarkovValidity}.
We note that for the current {[}Eq.\ (\ref{eqnCurrent}){]}, $\MeqSop$
and $\expectSopFull{l}$ are both evaluated at $\omega=0$, thus the
stationary current does not contain non-Markovian effects. However,
the Markov approximation changes the noise, which for $l=\lalt$ becomes
symmetrized, $S_{ll}^{\mathrm{Mkov}}(\omegans)=S_{ll}^{\mathrm{Mkov}}(-\omegans)$
\cite{BornMkovSymmetric}. In particular, the antisymmetric noise
contribution (showing pure quantum effects) cannot be obtained in
the Markov approximation.

We now return to the exact expression of the noise in Born approximation
{[}Eq. (\ref{SgeneralW}){]} and explicitly calculate the matrix elements
of the various superoperators, \begin{eqnarray}
\!-\!\memoryKernel(t)e^{it\LNot}\rhoSys\! & = & \!\sum_{l}\left(\calGPlusOutPlusIn\rhoSys-\LindGPlusOutPlusIn\,\rhoSys\right)+\mathrm{h.c.},\label{eq:memoryKernelcalG}\\
W_{l}^{<(>)}(t)e^{it\LNot} & = & (-1)^{l}e\Big(\stackrel{{\scriptscriptstyle (}\sim{\scriptscriptstyle )}}{\calGSymbol}\!\!{}_{-}^{l}\:\mp\;\LindGPlusOutMinusIn\Big),\\
\expectSopTwo{l}{\lalt}(t)e^{it\LNot} & = & \delta_{l\lalt}\: e^{2}\:\LindGPlusOutPlusIn,\label{eq:expectTwoCalG}\end{eqnarray}
 with $\calGPlusOutPMIn=\calGSymbol^{l,\,\tunnelOutDotLbl}(t)\pm\calGSymbol^{l,\,\tunnelInDotLbl}(t)$
and $\LindGPlusOutPMIn=\sum_{bn}\op{b}{n}\,\TrS\!\left\{ \,\calGPlusOutPMIn\op{n}{b}\,\right\} $
\cite{Tmap}. We define $\sqrtGammaLS{nm}=\sqrt{2\pi\nu_{l\sigma}}\,\sum_{p}t_{lp}^{\sigma}\bra{n}d_{p\sigma}\ket{m},$
with spin dependent density of states $\nu_{l\sigma}$ in lead $l$.
The matrix elements of the remaining superoperators are  \begin{eqnarray}
{}\!\!\calG{b}{c}{n}{m}{l,\,\tunnelInDotLbl}(\omega) & \!\!=\!\! & \sum_{\sigma}\frac{\sqrtGammaLS{mc}{\sqrtGammaLS{nb}}^{\!\!*}}{2}\left(f_{l}(\Delta_{bn}-\omega)+\frac{i\principalValueSym{bn}{+}}{\pi}\right),\label{eq:calGIn}\\
{}\!\!\calG{b}{c}{n}{m}{l,\,\tunnelOutDotLbl}(\omega) & \!\!=\!\! & \sum_{\sigma}\frac{{\sqrtGammaLS{cm}}^{\!\!\!*}\sqrtGammaLS{bn}}{2}\left(\!1-f_{l}(\Delta_{nb}+\omega)+\frac{i\principalValueSym{nb}{-}}{\pi}\!\right)\!,\quad\label{eq:calGOut}\end{eqnarray}
with $\Delta_{bn}=E_{b}-E_{n}$, and $\principalValueSym{bn}{\pm}=\log\{2\pi kT/[(1\mp1)\fermiUCutoff/2\pm\Delta_{bn}-\omega]\}+\mathrm{Re}\,\psi\!\left[\frac{1}{2}+i\left(\Delta_{bn}\mp\omega-\mu_{l}\right)/2\pi kT\right].$
Here, $\psi$ is the digamma function. The terms $\principalValueSym{bn}{\pm}$
arise from the  principal values $\mathcal{P}\int_{0}^{\infty}d\epsilon f_{l}(\epsilon)/(\epsilon-\Delta_{bn}+\omega)$
and $\mathcal{P}\int_{0}^{\fermiUCutoff}d\epsilon\left[1-f_{l}(\epsilon)\right]/(\Delta_{nb}-\epsilon+\omega)$
with bandwidth cutoff $\fermiUCutoff$. If we neglect $\omegans$
with respect to the large energies $\Delta_{bn}$ and $\fermiUCutoff-\Delta_{nb}$,
the first term of $\principalValueSym{}{\pm}$ (and thus $\fermiUCutoff$)
drops out in $S_{l\lalt}(\omegans)$. We note that the contribution
corresponding to Eq.\ (\ref{eq:expectTwoCalG}) has been calculated
for the symmetrized noise of a single electron transistor with a continuous
spectrum, using a phenomenological Langevin approach \cite{KoroktovLvin}.
With these results, Eqs.\ (\ref{SgeneralW}) and (\ref{eq:memoryKernelcalG})-(\ref{eq:calGOut}),
it is straightforward to find $S_{l\lalt}(\omegans)$ for an arbitrary
dot spectrum; one only needs to evaluate simple algebraic expressions. 

\begin{figure}
\centerline{\includegraphics[%
  width=65mm]{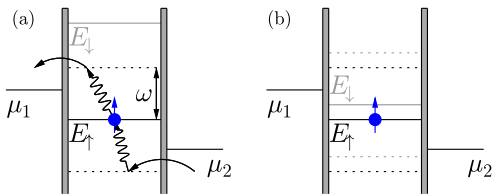}}

\caption{\label{cap:noiseDot}Quantum dot coupled to two leads and in the
sequential tunneling regime. (a) Large Zeeman splitting, $\Delta_{z}>\Dm+\omegans$.
When the dot is empty and $\omegans\geq\DEup-\mu_{2}$, an electron
from lead 2 absorbs energy $\omegans$ and tunnels for a short time
onto the dot, contributing to the noise $S_{22}$. Similarly, for
$\omegans>\mu_{1}-\DEup$, the electron on the dot can tunnel into
lead 1, contributing to $S_{11}$. (b) Smaller Zeeman splitting, here
$\Delta_{z}=\Dm/4$. When the energies $E_{\uparrow,\downarrow}\pm\omegans$
(dotted lines) are aligned with $\mu_{1,2}$, the shot noise has a
step, see Fig.\ \ref{cap:noiseF11-3lvl} (solid line).}
\end{figure}

We now identify the regime where the asymmetric noise properties become
most apparent. Asymmetries arise from the $\omegans$ dependence of
Eqs.\ (\ref{eq:calGIn})-(\ref{eq:calGOut}), i.e., are most prominent
for $|\omegans|>kT$, with steps occurring at $|\omega|\simeq|\Delta_{bn}-\mu_{l}|$
(see below). In this regime, the Markov approximation breaks down
\cite{MarkovValidity} and the noise probes non-Markovian effects.
Further, since the ST regime is valid for $kT>|\sqrtGammaLSSymbol|^{2}$,
we have $\omegans>|\sqrtGammaLSSymbol|^{2}$. Thus, the conditional
time evolution becomes $\condRho(\omega)\approx i/\omega$, since
$\MeqSop(\omega)\sim\sqrtGammaLSSymbol^{2}$. For the noise {[}Eq.~(\ref{SgeneralW}){]}
only the two last terms are relevant, since they are of order $\sqrtGammaLSSymbol^{2}$
while the other terms are of order $\sqrtGammaLSSymbol^{4}/\omega$
and can be neglected. Less formally, this is because multiple tunneling
processes {[}described by $\condRho(\omegans)${]} do not occur on
the short time scales corresponding to large $\omegans$. Thus, only
the individual (uncorrelated) tunneling events contribute, leading
to shot noise.

Next, we discuss specific cases, see Fig.\ \ref{cap:noiseDot}, where
a dot is coupled to two leads $l=1,2$ and a voltage bias $\Dm=\mu_{1}-\mu_{2}$
is applied. We assume single energy level spacing and Coulomb charging
energy larger than temperature, bias, and noise frequency. We consider
the dot states $\ket{0}$, with an even number of electrons and state
$\ket{\sigma}$ where an electron with spin $\sigma=\uparrow,\downarrow$
is added to the dot \cite{equivSingletSys}. For an applied magnetic
field $B$, the Zeeman splitting is $\Delta_{z}=g\mubohr B=\DEdown-\DEup>0$.
We consider the ST regime, $\mu_{1}>\DEup>\mu_{2}$, and define the
tunneling rates $\gls{l}{\sigma}=|\sqrtGammaLS{0\sigma}|^{2}$.

\emph{Dot with single level}. First, we assume a large Zeeman splitting
such that only the spin ground state $\spup$ is relevant, see Fig.\
\ref{cap:noiseDot}(a), and we omit the index $\uparrow$. Since in
this regime only one dot level is involved, there are no charging
effects between different levels. Thus, $\Hdot=\DEup d^{\dagger}d$
and so the full Hamiltonian $H$ is bilinear and can be solved exactly.
The symmetrized noise was calculated for this system and discussed
for $\omegans=0$ \cite{Averin93}. We now calculate the asymmetric
noise {[}Eq.\ (\ref{eqnNoiseDef}){]} for finite $\omegans$ exactly.
For this, we solve the Heisenberg equations for $d(t)$ and $c_{lk}(t)$
and find the current operator, $I_{l}(t)/e(-1)^{l}=\sum_{kl'k'}\left[\biResFactor{k'}\leadAnAlt^{\dagger}\leadAn+\mathrm{H.c.}\right]+\sum_{l'k'l''\kTwo}(\gl{l}/|t_{l}|^{2})\,\biResFactor{k'}\biResFactor{\kTwo}^{*}\leadAnAlt^{\dagger}\leadAnAltTwo$.
Here, the lead operators $\leadAn$ are evaluated at time $t_{0}$
(i.e., when the leads are at equilibrium) and we have defined $\biResFactor{k'}=i\,\tunnelAmp^{*}t_{l'}\, e^{i\left(\leadEnergyAlt-\leadEnergy\right)\left(t-t_{0}\right)}/\left(\leadEnergyAlt-\dotEnergy-i\levelWidth\right)$
and $\levelWidth=(\gO+\gT)/2$. Now we insert $I_{l}(t)$ into Eq.\
(\ref{eqnNoiseDef}) and readily obtain the asymmetric noise, containing
all quantum effects. We consider the coherent non-perturbative regime
of strong coupling to the leads in the quantum limit of large frequencies,
$\omegans>\levelWidth>kT$. We obtain the shot noise \begin{equation}
S_{ll}^{\lblTwoStates}(\omegans)=\sum_{l',\,\pm}\frac{\pm\gO\gl{l'}}{2\pi\levelWidth}\,\theta\left(\omega\pm\mu_{l'}\mp\mu_{l}\right)\left[h(\mu_{l'})-h(\mu_{l}\mp\omega)\right],\label{eq:Sll2lvlStrong}\end{equation}
where $h(\epsilon)=\arctan\left[(\epsilon-\DEup)/\levelWidth\right]$.
Note that the noise shows steps at $\omegans=\pm|\DEup-\mu_{l}|$
with width $\levelWidth$ \cite{thetaSteps}. Furthermore, for $\omegans>|\DEup-\mu_{l}|$,
$\Dm$, the noise is asymmetric and saturates at $S_{ll}^{\lblTwoStates}(\omega)=e^{2}\gl{l}$,
while $S_{ll}^{\lblTwoStates}(-\omega)=0$.

Let us now consider the ST regime $kT>\levelWidth$ in the exact solution.
For $\omegans>\levelWidth$, we then find  \begin{equation}
S_{ll}^{\lblTwoStates}(\omegans)=\sum_{l',\,\pm}\frac{\gl{l}\gl{l'}}{2\levelWidth}\left[\delta_{1,\mp1}\pm f_{l'}(\DEup)\right]\left[\delta_{1,\pm1}\mp f_{l}(\DEup\pm\omega)\right].\label{eq:Sll2lvlExact}\end{equation}
 Again, the noise shows a pronounced asymmetry. We can now compare
Eq.\ (\ref{eq:Sll2lvlExact}) with the noise obtained in the perturbative
approximation {[}Eq.\ (\ref{SgeneralW}){]} and find that they agree.
We further consider $\omegans>\Dm+kT$ in Eq.\ (\ref{eq:Sll2lvlExact})
such that $f_{l}(\DEup+\omegans)=0$ and $f_{l}(\DEup-\omegans)=1$,
leaving $f_{l}(\DEup)$ unrestricted. In this case, the (asymmetric)
shot noise is \begin{equation}
S_{ll}^{\lblTwoStates}(\omegans)=e^{2}\gl{l},\label{eq:Sll2Lvl}\end{equation}
 whereas $S_{ll}^{\lblTwoStates}(-\omegans)=S_{12}^{\lblTwoStates}(\pm\omegans)=0$;
this is the same result as we have found for strong coupling {[}Eq.\
(\ref{eq:Sll2lvlStrong}){]}. The interpretation is that for $S_{ll}^{\lblTwoStates}(-\omegans)$
the detector absorbs energy $\omegans$, which, however, cannot be
provided by any tunneling process. On the other hand, for $S_{ll}^{\lblTwoStates}(\omegans)$
the detector provides energy $\omegans$. Thus, if the dot is empty,
an electron with energy $\DEup-\omegans$ can tunnel from the Fermi
sea $l$ into the dot, and, if the dot is filled, an electron can
tunnel from the dot into an unoccupied lead state of energy $\DEup+\omega$,
see Fig.\ \ref{cap:noiseDot}(a) \cite{noisePAT}. In both cases,
the tunneling occurs with rate $\gl{l}$ and thus the contribution
to the autocorrelation is $e^{2}\gl{l}\,\delta(t)$. Note that for
$|\DEup-\mu_{l}|>kT$, the noise is $S_{11}^{\lblTwoStates}(\omegans)=e\expect{I}\,(\gO+\gT)/\gT$.
Thus, for large $\omegans$, the frequency dependent Fano factor,
$F_{11}(\omegans)=S_{11}(\omegans)/e\expect{I}$, is 2 for $\gO=\gT$,
and can even become larger for $\gO>\gT$, in contrast to the Markovian
case where we find it to be 1. Thus, we find that the quantum shot
noise is \emph{super-Poissonian}. Moreover, away from the ST regime,
say for $\DEup+kT>\mu_{l}$, the dot remains always in state $\ket{0}$
and only a small (higher-order in $H_{T}$) cotunneling current $\expect{I}$
flows through the dot \cite{NoiseLong}. However, the noise can still
be of lower order, it is $S_{ll}^{\lblTwoStates}(\omegans)=e^{2}\gl{l}f_{l}(\DEup-\omegans)$
for large $|\omegans|$, resulting in a large Fano factor $F_{ll}(\omegans)$
and super-Poissonian shot noise.

\begin{figure}
\includegraphics[%
  width=65mm]{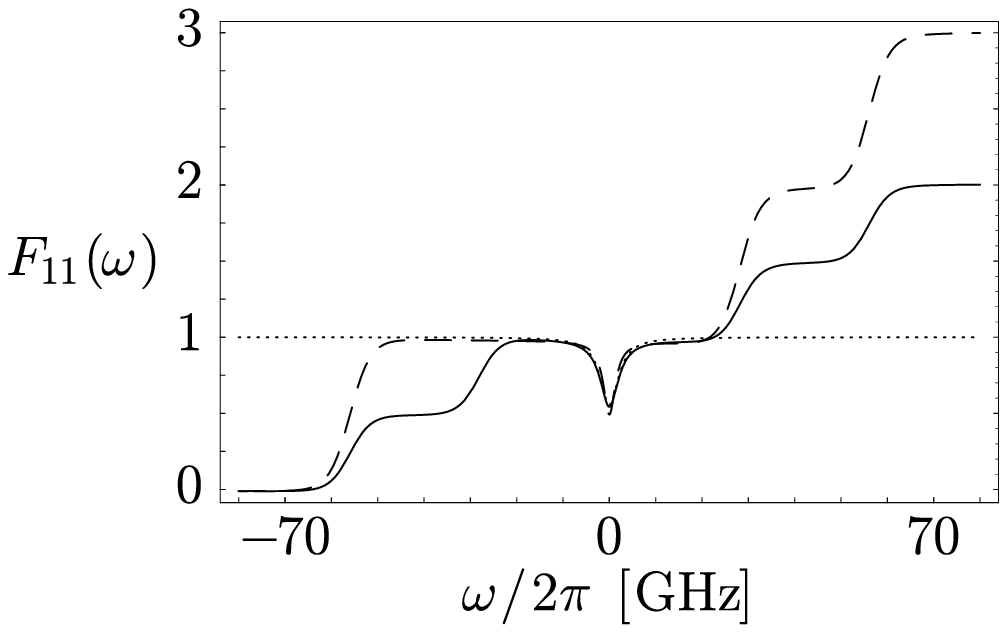}

\caption{\label{cap:noiseF11-3lvl}The Fano factor $F_{11}(\omegans)=S_{11}(\omegans)/e\expect{I}$
in the shot noise regime $\Dm>kT$ as function of noise frequency
$\omegans$. (Asymmetric noise at frequencies up to $90\,\mathrm{GHz}$
has been measured \cite{DeblockNoiseMeas}.) We consider $T=100\,\mathrm{mK}$,
$\Dm/e=460\,\mu V$, $\DEup=(\mu_{1}+\mu_{2})/2$, $\gO=\gT=5\times10^{9}\:\mathrm{s}^{-1}$,
and $g=2$. We use the full expression for the noise $S_{11}^{\mathrm{\lblThreeStatesSmZeeman}}$
{[}Eq.~(\ref{SgeneralW}){]} (solid line) and within Markov approximation,
$S_{11}^{\mathrm{\lblThreeStatesSmZeeman,Mkov}}$, (dotted line),
for $B=1\,\mathrm{T}$ {[}see Fig.~\ref{cap:noiseDot}(b){]}, thus
$\Delta_{z}=\Dm/4$ and $\expect{I}=530\,\mathrm{pA}$. We also show
$S_{ll}^{\lblThreeStatesImZeeman}$ (dashed line), being strongly
asymmetric, where $B=3\,\mathrm{T}$, $\Delta_{z}=3\Dm/4$, and $\expect{I}=400\,\mathrm{pA}$.
The dip near $\omegans=0$ is due to the charging effect of the dot,
while the steps at $\omegans_{i}$ (see text) arise from additional
transitions for increasing $\omegans$ and provide a striking effect
in the quantum shot noise. Note that these steps disappear when the
noise $S_{11}^{\mathrm{\lblThreeStatesSmZeeman}}$ is symmetrized
(some features remain for $\gO\neq\gT$). }
\end{figure}
\emph{Dot with two or more level}s. Second, we consider the regime
where the state $\spdown$ becomes relevant and charging interaction
enters (here no exact solution is available). We consider a small Zeeman splitting
such that $\mu_{1}>\DEupdown>\mu_{2}$ and $f_{l}(\DEup)\approx f_{l}(\DEdown)$,
see Fig.\ \ref{cap:noiseDot}(b). Using Eq.\ (\ref{SgeneralW}),
we calculate the noise $S_{11}^{\lblThreeStatesSmZeeman}(\omegans)$
and plot it in Fig.\ \ref{cap:noiseF11-3lvl} (solid line). For large
$|\omegans|$, such that $f_{l}(\DEnoise{\sigma}+|\omegans|)=0$ and
$f_{l}(\DEnoise{\sigma}-|\omegans|)=1$, the noise vanishes for $\omegans<0$
while for $\omegans>0$ it saturates at \begin{equation}
S_{ll}^{\lblThreeStatesSmZeeman}(\omegans)=2e^{2}\gl{l}\,\frac{\gO+\gT}{\gO[1+f_{1}(\DEup)]+\gT[1+f_{2}(\DEup)]}.\label{eq:Sll3lvlsmZeeman}\end{equation}
More generally, for the weaker assumption $|\omegans|>\levelWidth$,
the numerator in Eq.\ (\ref{eq:Sll3lvlsmZeeman}) becomes $\frac{1}{2}\sum_{l',\pm,\sigma}\gl{l'}\left[\delta_{1,\mp1}\pm f_{l'}(\DEup)\right]\left[\delta_{1,\pm1}\mp f_{l}(\DEnoise{\sigma}\pm\omega)\right]$.
Thus, the noise $S_{11}^{\lblThreeStatesSmZeeman}$ shows four steps
at $\omegans_{i}=\pm(\mu_{1}-\DEupdown)$, corresponding to the dotted
lines in Fig.\ \ref{cap:noiseDot}(b). The interpretation is that
for increasing $\omegans,$ more energy is available and more tunneling
processes are allowed. Namely, if $\omegans\geq-(\mu_{1}-\DEup)$,
an electron with spin $\uparrow$ from lead 1 can emit energy $\omegans$
and tunnel onto the dot, and if $\omegans\geq\mu_{1}-\DEup$, an electron
on the dot with spin $\uparrow$ can absorb energy $\omegans$ and
tunnel into lead 1. Analogous processes occur for spin $\downarrow$.
The steps in $S_{ll}^{\lblThreeStatesSmZeeman}$ are broadened due
to and temperature, and the step is $\propto\tanh\left[(\omegans-\omegans_{i})/2kT\right]$.
For $\Dm>kT,\,\Delta_{z}$, i.e., $f_{l}(\DEnoise{\sigma})=\delta_{l1}$,
the height of the steps in the Fano factor $F_{11}(\omega)$ is $\frac{1}{2}$
for $\omega_{i}<0$ and $\gO/2\gT$ for $\omega_{i}>0$. (These heights
change for spin-dependent tunneling, $\gul{l}\neq\gdl{l}$.)  Next,
we consider an intermediate Zeeman splitting, $\DEdown>\mu_{1}+kT$.
In this regime (c), the dot is either in state $\ket{0}$ or $\spup$,
while the state $\spup$ is never occupied and so no additional tunneling
process occurs for $\omegans\geq-(\DEdown-\mu_{1})$. Thus, the steps
in the noise are at $\omegans=\pm(\mu_{1}-\DEup)$ and at $\omegans=\DEdown-\mu_{1}$,
see Fig.~\ref{cap:noiseF11-3lvl} (dashed line). For large $\omegans$,
the noise saturates at \begin{equation}
S_{ll}^{\lblThreeStatesImZeeman}(\omegans)=e^{2}\sum_{l'=1,2}\gul{l'}\,\frac{\gul{l}+\gdl{l}[1-f_{l'}(\DEup)]}{\guO+\guT}.\label{eq:Sll3lvlImZeeman}\end{equation}
 Here, we have allowed for spin-dependent tunneling. If we exclude
the contributions involving state $\spdown$ by setting $\gdl{l}=0$,
we recover Eq.~(\ref{eq:Sll2Lvl}). Generally, we see that the shot
noise $S_{ll}(\omega)$ of a quantum dot consists of a series of steps
and is monotonically increasing, apart from features near $\omega=0$.
Each dot level with energy (chemical potential of the dot) $E_{j}$
gives rise to steps at $\pm(\mu_{l}-E_{j})$ if the level is inside
the bias window, $\mu_{1}+kT>E_{j}>\mu_{2}-kT$, and to a single step
at $|\mu_{l}-E_{j}|$ otherwise. We stress that such a highly asymmetric
$S_{ll}(\omega)$ can be observed with an appropriate measurement
apparatus \cite{Lesovik,Aguado,Imry}. For sufficiently large $|\omegans|$,
the antisymmetric contribution becomes $\frac{1}{2}\left[S_{ll}(\omegans)-S_{ll}(-\omegans)\right]=\frac{1}{2}\mathrm{sign}(\omega)S_{ll}(|\omegans|)$
and is, e.g., given  by Eqs.\ (\ref{eq:Sll2lvlExact})-(\ref{eq:Sll3lvlImZeeman}).

In conclusion, we have derived the asymmetric shot noise of a quantum
dot exactly for a single dot level and in the weak coupling regime
for many levels. We have shown that the shot noise exhibits strong
asymmetric and super-Poissonian effects in the quantum limit. We thank
F. Marquardt, W. Belzig, G. Burkard, A. Cottet, J.C. Egues, H. Gassman,
D. Saraga, and C. Schönenberger for discussions. We acknowledge support
from the Swiss NSF, NCCR Nanoscience Basel, DARPA, and ARO.

\clearpage

\begin{thebibliography}{10}
\bibitem{BlanterBuettiker}Ya.M. Blanter, M. Büttiker, Phys.\ Rep.\ \textbf{336}, 1 (2000).
\bibitem{Lesovik}G.B. Lesovik, R. Loosen, JETP Lett.\ \textbf{65} (3), 295 (1997).

\bibitem{Aguado}R. Aguado, L.P. Kouwenhoven, Phys.\ Rev.\ Lett.\ \textbf{84}, 1986
(2000).
\bibitem{Imry}U. Gavish, Y. Levinson, Y. Imry, Phys.\ Rev.\ B \textbf{62}, 10637
(2000).
\bibitem{DeblockNoiseMeas}R. Deblock \emph{et al.}, Science \textbf{301}, 203 (2003).
\bibitem{Birk}H. Birk, M.J.M. de Jong, C. Schönenberger, Phys.\ Rev.\ Lett.\
\textbf{75}, 1610 (1995).
\bibitem{Averin93}D.V. Averin, J. Appl.\ Phys.\ \textbf{73}, 2593 (1993).
\bibitem{Korotkov94}A.N. Korotkov, Phys.\ Rev.\ B \textbf{49}, 10381 (1994).
\bibitem{Hershfield}S. Hershfield \emph{et al.}, Phys.\ Rev.\ B \textbf{47}, 1967 (1993).
\bibitem{DDot}D. Loss, E.V. Sukhorukov, Phys.\ Rev.\ Lett.\ \textbf{84}, 1035
(2000).
\bibitem{NoiseLong}E.V. Sukhorukov, G. Burkard, D. Loss, Phys.\ Rev.\ B \textbf{63},
125315 (2001).
\bibitem{LandauLifshitzS118}L.D. Landau, E.M. Lifshitz, \emph{Statistical Physics},  Vol.\ 5,
\S118 (Pergamon, London/Paris, 1958).
\bibitem{kouwenhoven}L.P. Kouwenhoven, G. Sch\"{o}n, L.L. Sohn, in \textit{Mesoscopic
Electron Transport}, Vol.\  345 of \textit{NATO Advanced Study Institute,
Series E}, edited by L.L. Sohn, L.P. Kouwenhoven, G. Sch\"{o}n (Kluwer,
Dordrecht, 1997).
\bibitem{FickSauermann}E. Fick, G. Sauermann, \emph{The Quantum Statistics of Dynamic Processes}
(Springer, Berlin, 1990).
\bibitem{condRhoResum}The formal solution of Eq.\ (\ref{eqnMasterEq}) corresponds to resumming
an infinite number of terms. Since we evaluate the self-energy $\MeqSop$
in leading order, we only resum a certain subset of the higher order
terms in $\condRho(\omega)$.
\bibitem{MarkovValidity}Evaluating $\memoryKernel(\tau)$ explicitly, we find that it decays
on a time scale $\tauc\sim1/kT$, i.e., the correlations induced in
the leads decay within $\tauc$. Thus, the Markov approximation is
justified for $\omegans\tauc,\,\gAvg\tauc\ll1$.
\bibitem{BornMkovSymmetric} In
Born-Markov approximation, the autocorrelation function is always
symmetric in $\omegans$ for operators $\dot{X}$ with $\left[X,\, H_{0}\right]=\left[X,\,\rBNot\right]=0$. 
\bibitem{Tmap}This linear map $\calGSymbol\mapsto\LindSopToOpSymbol$ is the identity
on operators, however, here $\calGSymbol$ is a superoperator. With
this notation, Eq.~(\ref{eq:memoryKernelcalG}) is reminiscent of
the Lindblad form, $\sum_{i}[A_{i}\rhoSys A_{i}^{\dagger}-A_{i}^{\dagger}A_{i}\rhoSys]+\mathrm{h.c.}$
\bibitem{KoroktovLvin}A.N. Korotkov, Europhys.\ Lett.\ \textbf{43}, 343 (1998).
\bibitem{equivSingletSys}This system is equivalent to the one with an additional electron on
the dot and where the ground state with two electrons is a singlet
$\ket{S}$ \cite{ELesr}.
\bibitem{thetaSteps}The steps coming from the $\theta$-function are suppressed by the
last factor of Eq.~(\ref{eq:Sll2lvlStrong}).
\bibitem{noisePAT}This is similar to photon-assisted tunneling currents, where an oscillating
field provides or absorbs energy {[}C. Bruder, H. Schoeller, Phys.\
Rev.\ Lett.\ \textbf{72}, 1076 (1994){]}.
\bibitem{ELesr}H.-A. Engel, D. Loss, Phys.\ Rev.\ Lett.\ \textbf{86}, 4648 (2001);
Phys.\ Rev.\ B \textbf{65}, 195321 (2002). 
\end{thebibliography}
\end{document}